\documentclass[english,11pt,a4paper]{article}
\usepackage[T1]{fontenc}
\usepackage[latin9]{inputenc}
\usepackage{amsmath}
\usepackage{amsthm}
\usepackage{amssymb}
\usepackage{graphicx}

\makeatletter
\numberwithin{equation}{section}

\usepackage{jheppub}
\usepackage{orcidlink}
\allowdisplaybreaks
\usepackage{extarrows} 

\makeatother

\usepackage{babel}
\begin{document}
\title{High-frequency gravitational waves from axion inflation in
the weak-backreaction regime}

\abstract{Axion inflation, characterized by a Chern-Simons interaction
between the inflaton and a gauge field, provides a powerful mechanism
for generating primordial gravitational waves (GWs) through tachyonic
enhancement of the gauge field. While recent literature has predominantly
focused on the Strong Backreaction (SB) regime to maximize GW signals
for future interferometers, this regime suffers from computational
complexities as well as the risk of overproducing scalar perturbations.
In this work, we investigate gauge field amplification and GW production
strictly within the theoretically safer Weak Backreaction (WB) regime,
with a particular focus on the largely unexplored non-instantaneous
reheating phase. Because the tachyonic enhancement during slow-roll
typically increases as inflation approaches its end, it is crucial
to investigate how the production of GWs  behaves at the very end
of inflation and thereafter.  By continuously tracking the evolution
from slow-roll through reheating to radiation domination, we present
a complete picture of inflationary and post-inflationary GW production
in this framework. A particularly interesting feature of the post-inflationary
phase   is that the oscillatory behavior of the inflaton during reheating
leads to frequent sign-flips of the instability parameter $\xi$,
exciting both helical modes of the gauge field. Our analysis reveals
that axion inflation can naturally generate one of the strongest known
primordial GW signals at high-frequency bands. The yield is relevant
for future precision measurements of the effective number of neutrino
species, $N_{{\rm eff}}$, and also strongly motivates the development
of novel high-frequency GW detectors. 

}

\author[a]{Xun-Jie Xu \orcidlink{0000-0003-3181-1386}}
\author[a,b]{and Junyu Zhu \orcidlink{0009-0009-8345-9928}}
\affiliation[a]{Institute of High Energy Physics, Chinese Academy of Sciences, Beijing 100049, China}
\affiliation[b]{School of Physical Sciences, University of Chinese Academy of Sciences, Beijing 100049, China}
\emailAdd{xuxj@ihep.ac.cn} 
\emailAdd{zhujunyu@ihep.ac.cn} 
\preprint{\today}
\maketitle

\section{Introduction}

Cosmic inflation stands as the leading paradigm for the dynamics of
the early universe, successfully explaining the origin of the primordial
density perturbations that seeded cosmic structure. A fundamental
prediction of the standard slow-roll inflationary scenario is the
generation of a stochastic background of primordial gravitational
waves (GWs), characterized by a nearly scale-invariant tensor power
spectrum. Remarkably, the predicted amplitude of this primordial signal
lies close\footnote{See, e.g., Ref.~\cite{Caprini:2018mtu} for a comprehensive review
of  the theoretical predictions  and Ref.~\cite{Schmitz:2020syl}
for the sensitivity reach of future interferometer experiments. } to the sensitivity reach of proposed space-based GW interferometer
experiments, such as DECIGO~\cite{Seto:2001qf,Kudoh:2005as}, the
Big Bang Observer (BBO)~\cite{Crowder:2005nr,Corbin:2005ny}, and
$\mu$ARES~\cite{Sesana:2019vho}. 

While the standard single-field slow-roll paradigm offers a baseline
observational target, a compelling and open question in early-universe
cosmology is whether this primordial tensor spectrum can be significantly
modified by the presence of extra fields interacting with the inflaton.

A theoretically well-motivated  framework that addresses this question
is axion inflation~\cite{Garretson:1992vt,Anber:2006xt,Anber:2009ua},
 driven by an axion-like pseudo-scalar which  naturally couples to
a gauge field via a Chern-Simons term.  This framework  has been
extensively studied due to its rich phenomenology arising from the
axion-gauge interaction, including large non-Gaussianity~\cite{Barnaby:2010vf,Barnaby:2011qe,Barnaby:2011vw,Barnaby:2012tk,Barnaby:2012xt,Linde:2012bt,Meerburg:2012id,Bamba:2014vda},
magnetogenesis~\cite{Durrer:2010mq,Anber:2015yca,Fujita:2015iga,Adshead:2016iae,Gorbar:2021zlr,Sobol:2019xls,Sharma:2024nfu,Iarygina:2025ncl},
baryogenesis~\cite{Anber:2015yca,Jimenez:2017cdr,Domcke:2018eki,Domcke:2022kfs,Cado:2022pxk},
primordial black holes~\cite{Linde:2012bt,Bugaev:2013fya,Peloso:2016gqs,Garcia-Bellido:2016dkw,Domcke:2017fix,Ozsoy:2020kat,Domcke:2020zez,Unal:2023srk,Ozsoy:2023ryl,Domcke:2023tnn},
and primordial GWs~\cite{Cook:2011hg,Sorbo:2011rz,Barnaby:2011qe,Anber:2012du,Barnaby:2012xt,Bamba:2014vda,Ferreira:2014zia,Jimenez:2017cdr,Adshead:2018doq,Adshead:2019igv,Mirbabayi:2014jqa,Adshead:2019lbr,Dimastrogiovanni:2016fuu,Namba:2015gja,Domcke:2016bkh,Domcke:2019qmm,Garcia-Bellido:2016dkw,Garcia-Bellido:2023ser,Mukohyama:2014gba,Niu:2023bsr,Sharma:2024nfu,Corba:2024tfz,vonEckardstein:2025oic,vonEckardstein:2025elq}.
   As is well established in the literature,  when the inflaton
rolls down its potential, it triggers a tachyonic instability that
leads to exponential production of gauge field quanta, which  subsequently
act as a powerful secondary source of tensor perturbations, capable
of producing substantially enhanced, scale-dependent GW signals that
are highly relevant for the observational windows of upcoming GW interferometer
experiments.

Despite the theoretical appeal of this mechanism, recent studies (see,
e.g., \cite{vonEckardstein:2025oic}) have demonstrated that when
considering phenomenologically viable inflationary potentials, the
coupling strength required to generate GW signals within the reach
of future GW interferometers generally leads to Strong Backreaction
(SB) of the gauge field.  Axion inflation in the SB regime entails
considerable complexity due to its highly non-linear dynamics, which
necessitates complex lattice simulations to fully resolve the problem~\cite{Cuissa:2018oiw,Caravano:2021bfn,Caravano:2022epk,Figueroa:2024rkr,Sharma:2024nfu,Iarygina:2025ncl}.
Yet, driven by the desire to maximize the tensor yield,  much of
the recent literature has focused on the SB regime~\cite{Papageorgiou:2019ecb,Domcke:2020zez,Caravano:2021bfn,Figueroa:2023oxc,Peloso:2022ovc,Figueroa:2024rkr,Caravano:2022epk,Garcia-Bellido:2023ser}.
While the great effort made in the literature on investigating axion
inflation in the SB regime is certainly of significant importance,
it should be noted that, in addition to its computational complexity,
the SB regime may also pose  the issue of overproducing scalar perturbations,
which risk violating stringent observational constraints from the
cosmic microwave background and primordial black hole bounds~\cite{Linde:2012bt,Garcia-Bellido:2016dkw}.

In this work, we concentrate our analysis strictly within the Weak
Backreaction (WB)  regime, in which we believe many questions remain
largely unexplored. One of them concerns the compatibility of the
exponential growth of gauge field modes with a graceful exit from
inflation via non-instantaneous reheating.  As has been recognized
in the literature, the exponential growth  relies crucially on the
instability parameter $\xi\propto\dot{\phi}$ where $\dot{\phi}$
is the time derivative of the inflaton field. During the slow-roll
phase, $\dot{\phi}$  is relatively small and it generally increases\footnote{This is the case for most plateau models (e.g., the Starobinsky model,
$\alpha$-attractor T-models) currently favored by constraints on
the tensor-to-scalar ratio.} as inflation approaches  its end.  Consequently, the strongest tachyonic
enhancement typically occurs at the very end of inflation, where the
slow-roll conditions are violated and the inflaton velocity $\dot{\phi}$
sharply increases,  causing $\xi$ to peak dramatically. Since the
physical evolution of the fields does not abruptly cease at the inflationary
exit, a careful treatment of the subsequent reheating phase, in which
$\dot{\phi}$ can remain appreciably large and typically oscillates,
becomes necessary.

 To properly capture these dynamics, we perform a detailed numerical
and analytical analysis of the gauge-field production and the induced
tensor power spectrum, continuously evolving the system from the
slow-roll phase, through the reheating epoch, and eventually to radiation
domination. Whenever possible, we make comparison of our numerical
results with known analytical results obtained in previous studies.
While good agreement is often reached in such comparison when $\xi$
varies sufficiently slow, we also find large discrepancies when the
variation of $\xi$ becomes rapid, which typically occurs near the
end of inflation.  In particular, our analysis reveals that the range
of the SB regime may be overestimated if previous analytical formulae
for the gauge field energy density are used. 

Regarding the observational consequences, even though our analysis
confirms the conclusion of Ref.~\cite{vonEckardstein:2025oic} that
GW signals relevant for upcoming GW interferometers cannot be obtained
in the WB regime, we find that the WB regime still generates particularly
interesting high-frequency GWs. Due to the rapid growth of $\xi$
at the end of inflation and its subsequent oscillatory behavior, the
resulting high-frequency GW signal not only is many orders of magnitude
higher than the inflationary production, but also receives contributions
from both helical modes of the gauge field as a consequence of frequent
sign-flips of $\xi$ during reheating. The resulting high-frequency
GW yield is sufficiently high to be relevant to future precision measurements
of the effective number of neutrino species, $N_{{\rm eff}}$. Compared
to various studies on high-frequency GWs produced during reheating
from, e.g., inflaton decay~\cite{Barman:2023ymn,Barman:2023rpg,Bernal:2023wus,Kanemura:2023pnv,Tokareva:2023mrt,Barman:2024htg,Jiang:2024akb,Das:2025cqs,Bernal:2026dsu}
or particle scattering/annihilation~\cite{Choi:2024ilx,Xu:2024fjl,Bernal:2024jim,Bernal:2025lxp,Xu:2025wjq},
our calculation reveals that axion inflation may give rise to one
of the strongest primordial GW signals at high frequencies, thereby
motivating novel experimental developments in high-frequency GW detection,
which has emerged as a rapidly developing frontier in GW physics~\cite{Aggarwal:2020olq,Domcke:2023qle,Aggarwal:2025noe}.

The structure of this paper is organized as follows. In Sec.~\ref{sec:formulation},
we establish the framework for the background evolution covering inflation
and reheating. In Sec.~\ref{sec:partI}, we investigate the gauge
field amplification and the subsequent GW production during inflation.
This section also serves as a brief review of known analytical results,
against which we compare our numerical findings. In Sec.~\ref{sec:PartII},
we extend our analysis to the post-inflationary evolution, detailing
the complex dynamics of the system throughout the reheating epoch.
In Sec.~\ref{sec:hfgw}, we present our main observational results
regarding the high-frequency GW spectrum and discuss its implications
for $N_{{\rm eff}}$ and future high-frequency detectors. Finally,
we conclude in Sec.~\ref{sec:conclusion} and relegate some technical
details to the appendices.

\noindent {\bf Notation}: Throughout this work, we denote the physical
time by $t$, and the conformal time by $\eta$ or $\tau$. The derivatives
of a function $f$ with respect to $t$ and $\eta$ are denoted by
by $\dot{f}$ and $f'$, respectively. The Hubble parameter is $H=\dot{a}/a$
with $a$ the scale factor. Occasionally we  also use the conformal
Hubble parameter ${\cal H}\equiv a'/a=aH$. The FLRW metric is $ds^{2}=dt^{2}-a^{2}(t)d^{2}\mathbf{x}=a^{2}(\eta)\left(d\eta^{2}-d^{2}\mathbf{x}\right)$,
implying the $(+,-,-,-)$ metric signature.   We use the reduced
Planck mass $M_{P}=1/\sqrt{8\pi G}\simeq2.435\times10^{18}$ GeV.
The effective numbers of relativistic degrees of freedom in energy
and entropy are denoted by $g_{\star\rho}$ and $g_{\star s}$, respectively.

\section{Background evolution \label{sec:formulation}}

We start with an inflaton field, $\phi$, coupled to a gauge field,
$A^{\mu}$,  via a Chern-Simons term. The relevant part of the Lagrangian
reads:
\begin{equation}
{\cal L}\supset\frac{1}{2}\partial^{\mu}\phi\partial_{\mu}\phi-V(\phi)-\frac{1}{4}F^{\mu\nu}F_{\mu\nu}+\frac{1}{4}\frac{\phi}{\Lambda}F^{\mu\nu}\tilde{F}_{\mu\nu}\thinspace,\label{eq:-3}
\end{equation}
where $V(\phi)$ is the inflaton potential, $F^{\mu\nu}=\partial^{\mu}A^{\nu}-\partial^{\nu}A^{\mu}$
is the field strength tensor of the gauge field with $\tilde{F}_{\mu\nu}$
its dual form, and $\Lambda$ is a dimensional constant to quantify
the coupling strength. The inflaton potential $V(\phi)$ in principle
can take a variety of forms and among them, many plateau models generally
remain phenomenologically viable. For concreteness, we take the $\alpha$-attractor
T model for a case study~\cite{Kallosh:2013hoa}:

\begin{equation}
V(\phi)=\lambda M_{P}^{4}\left[\sqrt{6}\tanh\left(\frac{\phi}{\sqrt{6}M_{P}}\right)\right]^{2}.\label{eq:-13}
\end{equation}
It is known that due to its flatness at large $\phi$, the potential
leads to a small but detectable tensor-to-scalar ratio $r=12/N_{e}^{2}$
with $N_{e}\approx55$ the number of e-folds for CMB observations.
The dimensionless coefficient $\lambda$ can be determined by $\lambda\approx3\pi^{2}A_{s}/N_{e}^{2}\approx2.055\times10^{-11}$~\cite{Barman:2022qgt}
where $A_{s}\approx2.1\times10^{-9}$~\cite{Planck:2018jri} is the
amplitude of scalar fluctuations. The potential has a quadratic bottom:
\begin{equation}
V\approx\frac{1}{2}m_{\phi}^{2}\phi^{2}\ \ \text{with}\ \ m_{\phi}^{2}\equiv2\lambda M_{P}^{2}\thinspace,\label{eq:-52}
\end{equation}
implying that when $\phi$ completes the slow-roll evolution and starts
oscillating around the minimum, it behaves as matter. Physically,
it can be interpreted as a cold, dense condensate of inflaton particles
with mass $m_{\phi}$. In the subsequent evolution, reheating can
be achieved by assuming that these inflaton particles gradually decay
to radiation with the decay rate denoted by $\Gamma_{\phi}$. 

With the above setup, the background evolution is governed by the
following equations:
\begin{align}
\ddot{\phi}+(3H+\Gamma_{\phi})\dot{\phi}+\frac{dV}{d\phi} & =\frac{\rho_{EB}}{\Lambda}\thinspace,\label{eq:-4}\\
\dot{\rho}_{R}+4H\rho_{R} & =\Gamma_{\phi}\left(\rho_{\phi}+p_{\phi}\right),\label{eq:-5}
\end{align}
 where $H=\dot{a}/a$ is the Hubble parameter with $a$ the scale
factor, $\rho_{R}$ is the energy density of radiation, and $\rho_{\phi}=\frac{1}{2}\dot{\phi}^{2}+V$
and $p_{\phi}=\frac{1}{2}\dot{\phi}^{2}-V$ are the energy density
and pressure of $\phi$, respectively.  In Eq.~\eqref{eq:-4}, $\rho_{EB}\equiv\langle\mathbf{E}\cdot\mathbf{B}\rangle$
is the ensemble average of $\mathbf{E}\cdot\mathbf{B}$ with $\mathbf{E}_{i}\equiv F_{0i}/a^{2}$
and $\mathbf{B}_{i}\equiv-\epsilon_{ijk}F^{jk}/a^{2}$. It accounts
for the backreaction of the gauge field to the evolution of the inflaton
field. In the WB regime, its influence is negligibly small. We will
discuss the backreaction in later sections. 

\begin{figure}
\centering

\includegraphics[width=0.49\textwidth]{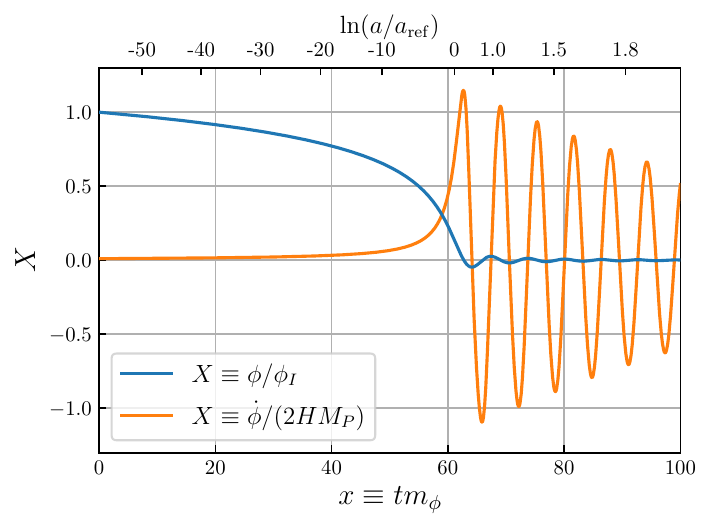}\includegraphics[width=0.49\textwidth]{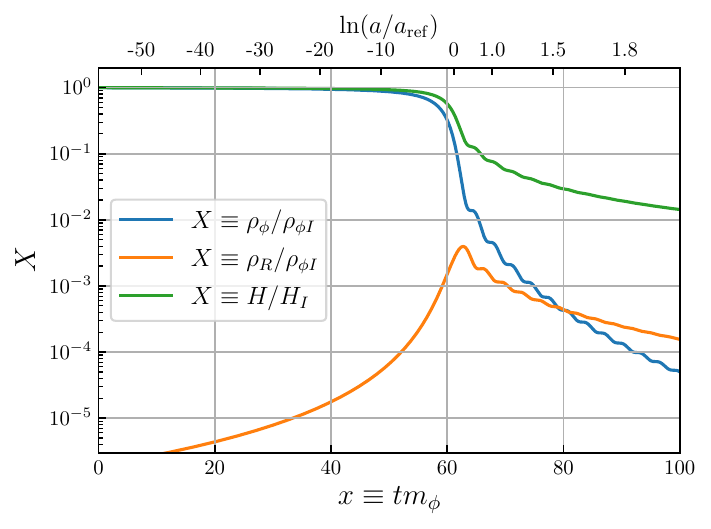}\caption{Left panel: the evolution of $\phi$ and $\dot{\phi}$ from inflation
to reheating. Right panel: the evolution of $H$, $\rho_{\phi}$,
and $\rho_{R}$ during the same epoch. \label{fig:bkg}}
\end{figure}

 In Fig.~\ref{fig:bkg}, we present the numerical solutions of Eqs.~\eqref{eq:-4}
and \eqref{eq:-5} with $\Gamma_{\phi}=0.05m_{\phi}$, $m_{\phi}=6.41\times10^{-6}M_{P}$
(determined from $\lambda$), and $\Lambda^{-1}=0$. We start the
evolution with the initial value of $\phi$ set at $\phi_{I}=-6.13M_{P}$,
corresponding to $N_{e}=55$. The top x-axis indicates the value of
$\ln(a/a_{{\rm ref}})$ where $a_{{\rm ref}}$ is the scale factor
when ${\cal H}\equiv aH$ reaches the maximum, i.e., 
\begin{equation}
a_{{\rm ref}}H(a_{{\rm ref}})={\cal H}_{{\rm max}}\thinspace.\label{eq:-58}
\end{equation}
 Since ${\cal H}$ increases during inflation and decreases during
matter or radiation domination, $a_{{\rm ref}}$ can be approximately
regarded as the scale factor at the end of inflation, implying 
\begin{equation}
-\ln(a/a_{{\rm ref}})\approx N_{e}\thinspace.\label{eq:-73}
\end{equation}
The bottom x-axis indicates the value of $x\equiv tm_{\phi}$ where
$t$ is the physical time. Since a large part of our analysis will
be dedicated to the post-inflationary epoch where $\phi$ oscillates
periodically with the frequency $\omega=m_{\phi}$, it is convenient
to use $x$ instead of $t$ as a dimensionless time variable in this
work. 

From Fig.~\ref{fig:bkg}, we see that for $x\lesssim60$, the evolution
is inflationary, meaning that $H$ is approximately constant and $a$
increases exponentially with $t$. For $x\gtrsim60$, $\rho_{\phi}$
decreases significantly due to inflaton decay and Hubble expansion.
Eventually the universe becomes radiation dominated after $\rho_{R}$
exceeds $\rho_{\phi}$ at $x\approx80$. For later convenience, we
define the reheating scale factor $a_{{\rm rh}}$ by
\begin{equation}
\rho_{R}(a_{{\rm rh}})=\rho_{\phi}(a_{{\rm rh}})\thinspace.\label{eq:-53}
\end{equation}
For the shown example, $a_{{\rm rh}}/a_{{\rm ref}}\approx\exp(1.6)\approx5$,
i.e., the universe expands by roughly a factor of $5$ from the end
of inflation to the completion of reheating. 

\section{Gauge field and GW production during inflation \label{sec:partI}}

\subsection{Gauge field production}

 The Chern-Simons term in axion inflation is known to be capable
of developing tachyonic instability of the gauge field during inflation,
thereby exponentially amplifying one of its helical modes~\cite{Cook:2011hg}.
Below we briefly review the formalism of the gauge field production
in axion inflation and discuss the tachyonic instability. 

First, let us quantize the gauge field:
\begin{equation}
A^{\mu}(\eta,\ \mathbf{x})=\int\frac{d^{3}\mathbf{k}}{(2\pi)^{3}}\sum_{\lambda=\pm}\left[\epsilon_{\lambda}^{\mu}A_{\lambda,\mathbf{k}}(\eta)e^{i\mathbf{k}\cdot\mathbf{x}}a_{\lambda,\mathbf{k}}+\epsilon_{\lambda}^{*\mu}A_{\lambda,\mathbf{k}}^{*}(\eta)e^{-i\mathbf{k}\cdot\mathbf{x}}a_{\lambda,\mathbf{k}}^{\dagger}\right],\label{eq:-55}
\end{equation}
where $\epsilon_{\lambda}^{\mu}$ is the polarization vector, $A_{\lambda,\mathbf{k}}(\eta)$
is the mode function, and $a_{\lambda,\mathbf{k}}$ and $a_{\lambda,\mathbf{k}}^{\dagger}$
are the annihilation and creation operators satisfying $[a_{\lambda,\mathbf{k}},\thinspace a_{\lambda',\mathbf{k'}}^{\dagger}]=(2\pi)^{3}\delta_{\lambda\lambda'}\delta^{(3)}(\mathbf{k}-\mathbf{k}')$.\footnote{Here we adopt the convention of Ref.~\cite{Peskin:1995ev} for field
quantization. This causes some normalization differences compared
to, e.g., Refs.~\cite{Jimenez:2017cdr,vonEckardstein:2025oic}, where
$\int\frac{d^{3}\mathbf{k}}{(2\pi)^{3/2}}$ instead of $\int\frac{d^{3}\mathbf{k}}{(2\pi)^{3}}$
is used and a factor of $(2\pi)^{3}$ is removed from the commutation
relation of $a_{\lambda,\mathbf{k}}$ and $a_{\lambda,\mathbf{k}}^{\dagger}$.
The other parts including the mode function are not affected.} For simplicity, we will suppress the subscript $\mathbf{k}$ of $A_{\lambda,\mathbf{k}}(\eta)$
unless the momentum dependence needs to be emphasized. 

Note that here we use the conformal time $\eta$ instead of the physical
time $t$ such that the spacetime is locally Minkowski-like, i.e.,
$ds^{2}=a^{2}(\eta)\left(d\eta^{2}-d^{2}\mathbf{x}\right)$, which
allows for canonical quantization in a familiar form.  Globally,
since the time translation invariance is broken, the mode function
$A_{\lambda}$ does not evolve simply as $e^{-ik\eta}$ (the form
that would appear  in Minkowski QFT) and requires solving the corresponding
mode equation. 

In general, the mode function is normalized by the Wronskian condition:
\begin{equation}
A_{\lambda}{A'}_{\lambda}^{*}-A'_{\lambda}A{}_{\lambda}^{*}=i\thinspace,\label{eq:-56}
\end{equation}
with the initial conditions determined by the  Bunch-Davies vacuum:
\begin{equation}
A_{\lambda}\to\frac{1}{\sqrt{2k}}e^{-ik\eta}\ \ \text{and}\ \ A'_{\lambda}\to-ikA_{\lambda}\ \ \text{for}\ \ \eta\to-\infty\thinspace.\label{eq:-9}
\end{equation}

In the presence of the Chern--Simons term,  the equation of motion
for the mode function is given by~\cite{Cook:2011hg}:
\begin{align}
A_{\lambda}''+(k^{2}-2\lambda\xi k{\cal H})A_{\lambda} & =0\thinspace,\label{eq:-6}
\end{align}
where
\begin{equation}
\xi\equiv\frac{\dot{\phi}}{2H\Lambda}\thinspace.\label{eq:-7}
\end{equation}
During slow-roll inflation, $\xi$ varies mildly while ${\cal H}=aH$
increases rapidly due to the exponentially growth of $a$. When $2|\xi|{\cal H}$
exceeds $k$, Eq.~\eqref{eq:-6} implies that the mode with positive
$\lambda\xi$ becomes tachyonic. Consequently, the mode function grows
exponentially until the mode exits the horizon, which happens when
the physical momentum $k_{{\rm phy}}\equiv k/a$ becomes smaller than
$H$. 

Combining the two qualitative conditions, $2|\xi|{\cal H}\gtrsim k$
and $k_{{\rm phy}}\gtrsim H$,  one expects that $A_{\lambda}$ with
a proper sign of $\lambda$ grows substantially during 
\begin{equation}
{\cal H}\lesssim k\lesssim2|\xi|{\cal H}\thinspace.\label{eq:-8}
\end{equation}
In the slow-roll epoch, where ${\cal H}\approx-\frac{1}{\eta}$ and
$\eta<0$, Eq.~\eqref{eq:-8} is equivalent to 
\begin{equation}
1\lesssim-k\eta\lesssim2|\xi|\thinspace,\label{eq:-57}
\end{equation}
which can be viewed as a qualitative estimate of the time window for
the exponential growth.

Quantitatively, one can solve Eq.~\eqref{eq:-6} to examine the growth
of $A_{\lambda}$, which can be done both analytically or numerically. 

The analytical solution, under the assumption of exact de Sitter space
and constant $\xi$, can be found in Ref.~\cite{Jimenez:2017cdr}:
\begin{equation}
A_{\lambda}=\frac{e^{\lambda\pi\xi/2}}{\sqrt{2k}}W_{-i\lambda\xi,1/2}\left(2ik\eta\right),\label{eq:-10}
\end{equation}
where $W$ is the Whittaker function. For $k|\eta|\gg1$, Eq.~\eqref{eq:-10}
reduces to Eq.~\eqref{eq:-9}. For $k|\eta|\ll1$, Eq.~\eqref{eq:-10}
reduces to the following limit:
\begin{equation}
\lim_{\eta\to0}A_{\lambda}=\frac{e^{\pi\lambda\xi/2}}{\sqrt{2k}}\cdot\frac{1}{i\lambda\xi\Gamma(i\lambda\xi)}\thinspace,\label{eq:-11}
\end{equation}
where $\Gamma$ is the Euler gamma function. At large $\xi$, it is
a good approximation to use Stirling's formula: $\Gamma(i\lambda\xi)\approx\sqrt{2\pi/|\xi|}e^{-\pi|\xi|/2}e^{i\lambda\left(|\xi|\log|\xi|-|\xi|-\frac{\pi}{4}\right)}$,
which implies that Eq.~\eqref{eq:-11} is exponentially amplified
by $e^{\pi\lambda\xi/2}e^{\pi|\xi|/2}=e^{\pi|\xi|}$ for ${\rm sign}(\lambda\xi)=+$
(either $\lambda=+$ with $\xi>0$, or $\lambda=-$ with $\xi<0$).
If one flips the sign of $\lambda\xi$, the two exponential factors
cancel out, leading to no enhancement of the mode function. Without
loss of generality, we assume that $\xi$ during slow-roll is positive
throughout this work. Then only the $\lambda=+$ mode  is amplified
during slow-roll\footnote{Note that after slow-roll, $\xi\propto\dot{\phi}$ may frequently
flip its sign when $\phi$ oscillates around the minium of the potential.
Consequently, the $\lambda=-$ mode can also be amplified during reheating---see
Fig.~\ref{fig:high-k} for illustration. }. The amplification factor is approximately given by
\begin{equation}
\sqrt{2k}|A_{+}|\approx\begin{cases}
\frac{e^{\pi\xi}}{\sqrt{2\pi\xi}} & \xi\gtrsim0.2\\
1+\frac{\pi}{2}\xi & 0<\xi\lesssim0.2
\end{cases}\ \ (\text{for}\ k|\eta|\ll1)\thinspace,\label{eq:-15}
\end{equation}
with a maximal error of about $18\%$ occurring at $\xi=0.2$. Note
that Eq.~\eqref{eq:-15} should only be applied to super-horizon modes
and, if the variation of $\xi$ is non-negligible, $\xi$ should be
evaluated approximately at horizon crossing. 

\begin{figure}
\centering

\includegraphics[width=0.7\textwidth]{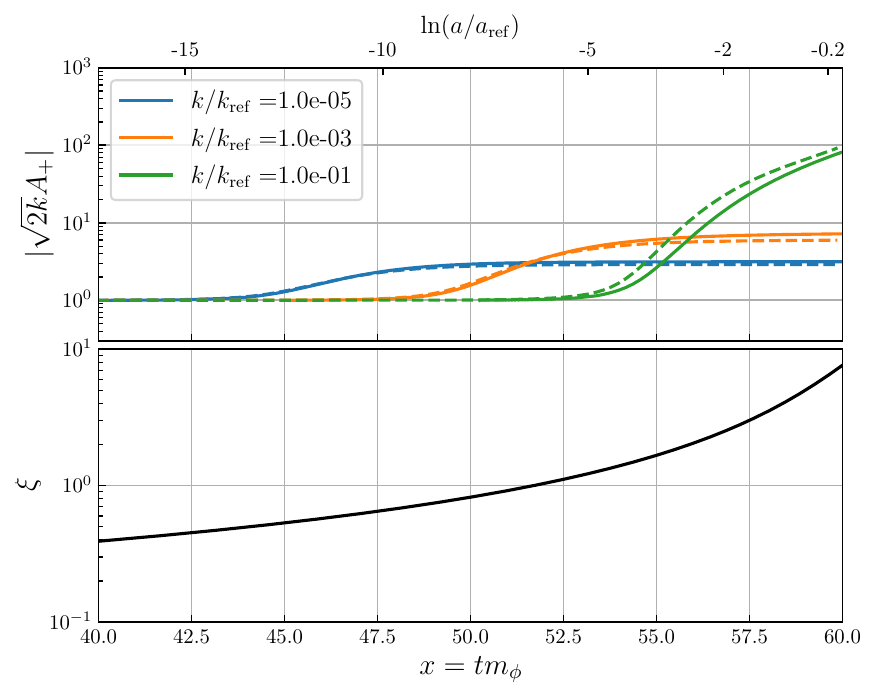}\caption{Upper panel: the evolution of the mode function $A_{+}$ during the
slow-roll epoch, with dashed and solid lines representing the analytical
and numerical solutions. Lower panel: the evolution of $\xi$ during
the same epoch. In this figure, we assume $\Lambda^{-1}=11.2/M_{P}$
and $\Gamma_{\phi}=0$. \label{fig:A-sol}}
\end{figure}

The numerical solution can be obtained by solving Eq.~\eqref{eq:-6}
straightforwardly.  In the upper panel of Fig.~\ref{fig:A-sol},
we show both numerical and analytical solutions for $k/k_{{\rm ref}}\in\{10^{-5},\ 10^{-3},\ 10^{-1}\}$
where $k_{{\rm ref}}\equiv a_{{\rm ref}}H(a_{{\rm ref}})={\cal H}_{{\rm max}}$.
 In the lower panel, we show the corresponding values of $\xi$. 

Note that the analytical solutions, derived under the assumption of
constant $\xi$, are exponentially sensitive to $\xi$. Although the
variation of $\xi$ during slow-roll is much  milder than $\eta$,
its mild variation nevertheless leads to different levels of enhancement
of $|A_{\lambda}|$. Hence for each analytical solution presented
here, one needs to choose an appropriate value of $\xi$ which should
be representative within the time interval in Eq.~\eqref{eq:-57}.
In Fig.~\ref{fig:A-sol}, this value is set at horizon crossing;
and the resulting analytical solutions reach modest agreement with
the numerical ones. 

One might consider incorporating the variation of $\xi$ into Eq.~\eqref{eq:-10}
by replacing the constant $\xi$ with the $\xi$ curve shown in the
lower panel of Fig.~\ref{fig:A-sol}. However, this would lead to
large deviations of the mode functions from their true values, as
we have  verified numerically.

\subsection{Gauge-field-induced GWs}

The exponentially enhanced gauge field can induce GW signals sufficiently
strong to be relevant for GW interferometer experiments.   The gauge-field-induced
GWs are typically calculated from the induced tenser power spectrum,
${\cal P}_{T}^{{\rm ind}}$, which contains two polarized contributions
(see, e.g., Ref.~\cite{vonEckardstein:2025oic}):
\begin{equation}
{\cal P}_{T}^{{\rm ind}}=\frac{1}{2}\sum_{\lambda=\pm}{\cal P}_{T,\lambda}^{{\rm ind}}\thinspace,\label{eq:-16}
\end{equation}
where 
\begin{align}
{\cal P}_{T,\lambda}^{{\rm ind}}\left(\eta,k\right) & =\frac{k^{3}}{2\pi^{2}M_{P}^{4}}\int\frac{d^{3}\mathbf{p}}{(2\pi)^{3}}\sum_{\lambda_{1}\lambda_{2}}\left(1+\lambda\lambda_{1}\frac{\mathbf{k}\cdot\mathbf{p}}{kp}\right)^{2}\left(1+\lambda\lambda_{2}\frac{\mathbf{k}\cdot\mathbf{q}}{kq}\right)^{2}\left|I_{\tau}(p,q)\right|^{2},\label{eq:-17}\\
I_{\tau}(p,q) & =\int_{-\infty}^{\eta}d\tau\frac{G_{k}(\eta,\tau)}{a(\tau)^{2}}\left[A'_{\lambda_{1},p}(\tau)A'_{\lambda_{2},q}(\tau)+\lambda_{1}\lambda_{2}pqA{}_{\lambda_{1},p}(\tau)A{}_{\lambda_{2},q}(\tau)\right].\label{eq:-18}
\end{align}
The momenta $\mathbf{p}$ and $\mathbf{q}$ satisfy $\mathbf{p}+\mathbf{q}=\mathbf{k}$. 

Eq.~\eqref{eq:-18} involves the retarded Green function $G_{k}(\eta,\tau)$,
which in exact de Sitter space has an analytical form~\cite{Cook:2011hg}:
\begin{equation}
G_{k}(\eta,\tau)\xlongequal{\text{de Sitter}}\frac{1}{k^{3}\tau^{2}}\left[\left(1+k^{2}\eta\tau\right)\sin(k(\eta-\tau))-k(\eta-\tau)\cos(k(\eta-\tau))\right]\theta\left(\eta-\tau\right).\label{eq:-29}
\end{equation}
For a more general spacetime background, the Green function will be
calculated in Sec.~\ref{subsec:The-Green-function}. 

Note that ${\cal P}_{T}^{{\rm ind}}$ should be added on top of the
tenser power spectrum arising from vacuum fluctuations of the metric,
denoted by ${\cal P}_{T}^{\text{vac}}$, which is known to be nearly
scale-invariant. Hence the combined tenser power spectrum is given
by
\begin{equation}
{\cal P}_{T}={\cal P}_{T}^{\text{vac}}+{\cal P}_{T}^{{\rm ind}},\ \ \text{with}\ \ {\cal P}_{T}^{\text{vac}}=\frac{2}{\pi^{2}}\left(\frac{H}{M_{P}}\right)^{2}\thinspace.\label{eq:-21}
\end{equation}

In Ref.~\cite{Cook:2011hg}, an analytical expression for ${\cal P}_{T,\lambda}^{{\rm ind}}$
has been derived, which reads:
\begin{equation}
{\cal P}_{T,\pm}^{{\rm ind}}\left(\eta,k\right)=2\frac{H^{4}}{\pi^{2}M_{P}^{4}}\frac{e^{4\pi\xi}}{\xi^{6}}f_{\pm}(\xi)\thinspace,\label{eq:-19}
\end{equation}
where $f_{\pm}$ are two dimensionless coefficients with weak dependence
on $\xi$. The original paper obtained $f_{+}=8.6\times10^{-7}$ and
$f_{-}=1.8\times10^{-9}$ \cite{Cook:2011hg}. In our numerical calculation,
we find  $f_{+}/10^{-7}\approx6.5+0.95\ln\left(\xi-1\right)$ and
$f_{-}/10^{-9}\approx2.3-0.2\ln\left(\xi-1\right)$ for $\xi\in[2,\ 10]$,
consistent with the values obtained in Ref.~\cite{Cook:2011hg}.

\subsection{Conditions for the WB regime}

\begin{figure}
\centering

\includegraphics[width=0.49\textwidth]{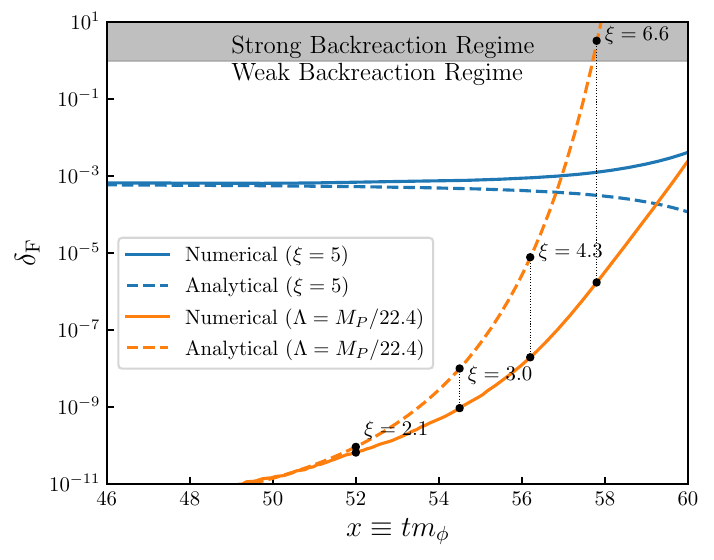}\includegraphics[width=0.49\textwidth]{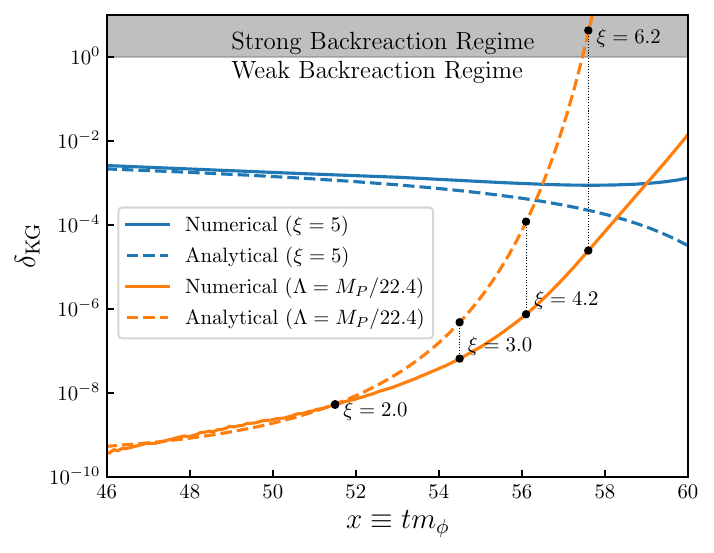}\caption{Analytical and numerical values of $\delta_{\text{F}}$ and $\delta_{\text{KG}}$
for two cases: (i) $\xi$ is fixed at a constant during slow-roll
and (ii) $\xi$ is determined by $\xi=\dot{\phi}/(2H\Lambda)$ with
$\Lambda=M_{P}/22.4$ and $\Gamma_{\phi}=0$. \label{fig:delta}}
\end{figure}

To identify the WB regime in which the backreaction of the gauge field
can be neglected,  it is useful to introduce the following two backreaction
parameters $\delta_{\text{F}}$ and $\delta_{\text{KG}}$~\cite{Jimenez:2017cdr,vonEckardstein:2025oic}:
\begin{equation}
\delta_{\text{F}}\equiv\frac{\rho_{A}}{\rho_{{\rm tot}}}\thinspace,\ \ \delta_{\text{KG}}\equiv\left|\frac{\rho_{EB}/\Lambda}{3H\dot{\phi}}\right|\thinspace.\label{eq:-22}
\end{equation}
where $\rho_{A}$ and $\rho_{{\rm tot}}=\rho_{\phi}+\rho_{R}$ denote
the gauge field and total energy densities. The WB regime requires
$\delta_{{\rm F}}\ll1$ and $\delta_{\text{KG}}\ll1$.  

Using analytical solutions of $A_{\lambda}$, one can estimate the
two parameters as follows~\cite{Jimenez:2017cdr,vonEckardstein:2025oic}:
\begin{align}
\delta_{\text{F}} & \approx4.1\times10^{-5}\left(\frac{H}{M_{P}}\right)^{2}\frac{e^{2\pi|\xi|}}{|\xi|^{5}}\left(1+1.1|\xi|^{2}\right),\label{eq:-23}\\
\delta_{\text{KG}} & \approx8.7\times10^{-5}\frac{H^{3}}{\dot{\phi}\Lambda}\frac{e^{2\pi|\xi|}}{|\xi|^{4}}\thinspace,\label{eq:-24}
\end{align}
which exhibit good  accuracy if $\xi$ is fixed at a constant above
${\cal O}(1)$ and if the cosmic expansion is approximately de Sitter---see
the blues curves in Fig.~\ref{fig:delta}. In practice, however,
Eqs.~\eqref{eq:-23} and \eqref{eq:-24} may lead to an overestimate
of $\delta_{\text{F}}$ and $\delta_{\text{KG}}$, as is illustrated
by the orange curves which are produced with $\xi$ determined by
the slow-roll dynamics and $\Lambda=M_{P}/22.4$. For instance, when
$\xi$ increases to $6.6$, Eq.~\eqref{eq:-23} leads to $\delta_{\text{F}}>1$,
 implying that the system enters the strong backreaction regime,
while numerical results suggest that it is still well in the WB regime.
Therefore, for the $\alpha$-attractor T model with $\Lambda=M_{P}/22.4$,
the inflationary part ($x\lesssim60$) of the evolution is in the
WB regime. In the next section, we will examine whether it will remain
in the WB regime during the reheating epoch.

\section{Post-inflationary evolution \label{sec:PartII}}

In this section, we will extend the previous analysis into the post-inflationary
epoch. The post-inflationary evolution features large oscillatory
$\xi$, which may not only amplify the gauge field mode with $\lambda=+$
substantially, but also lead to significant enhancement of the opposite-helicity
mode. Consequently, we expect rich phenomenology arising from the
post-inflationary epoch. 

\subsection{High-$k$ modes of the gauge field}

\begin{figure}
\centering

\includegraphics[width=0.99\textwidth]{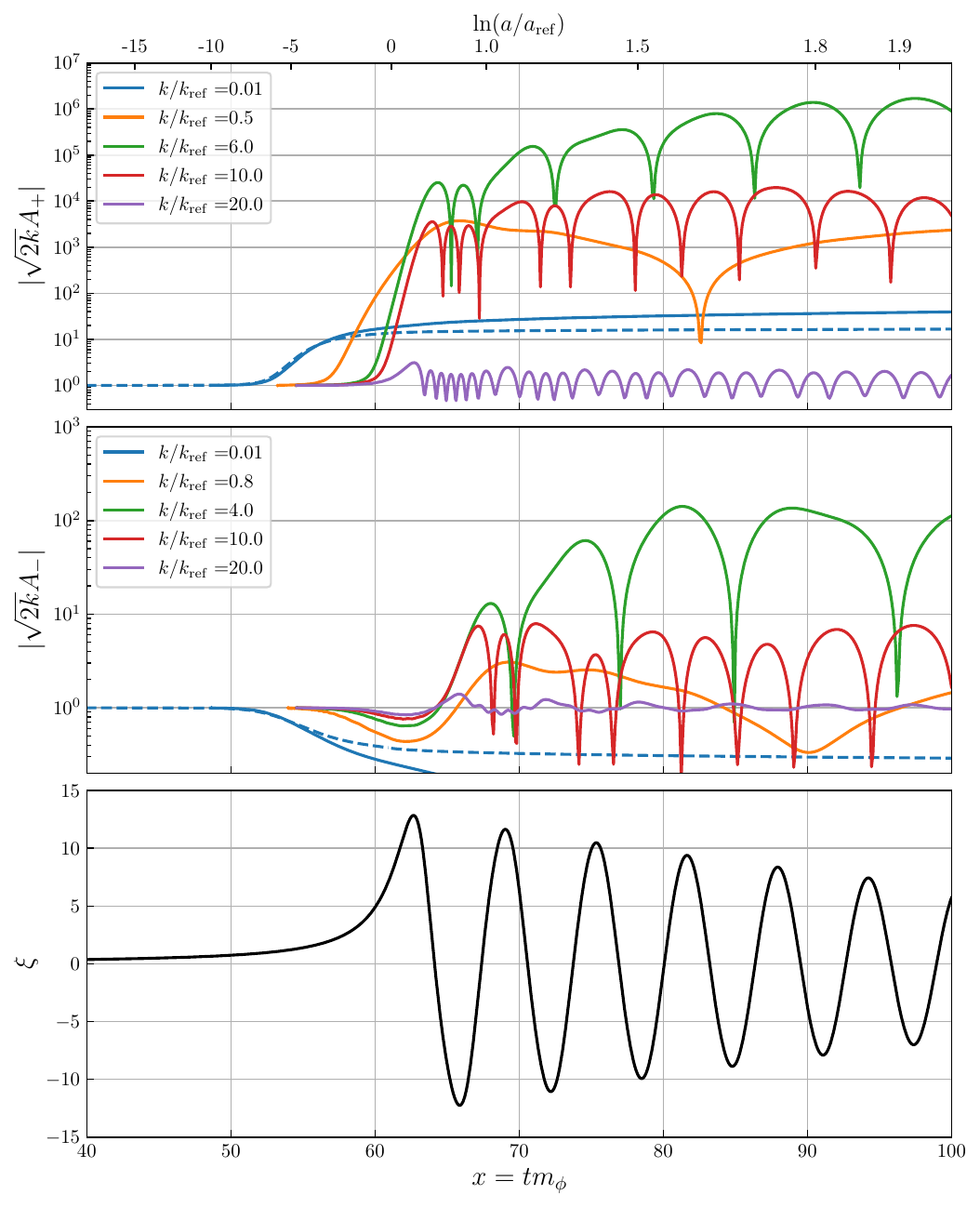}

\caption{The evolution of high-$k$ modes spanning inflationary ($tm_{\phi}\lesssim60$)
and reheating epochs ($tm_{\phi}\gtrsim60$). The upper and middle
panels are for $A_{+}$ and $A_{-}$, respectively. The lower  panel
shows the corresponding value of $\xi$. This figure is produced with
$\Lambda^{-1}=11.2/M_{P}$ and $\Gamma_{\phi}=0.05m_{\phi}$. \label{fig:high-k}
The blue dashed line represent the analytical estimate using the Whittaker
function {[}see Eq.~\eqref{eq:-10}{]}, which is valid only for $k\ll k_{{\rm ref}}$.
}
\end{figure}

As already shown in Fig.~\ref{fig:A-sol}, high-$k$ modes of the
gauge field are amplified more significantly than low-$k$ modes.
This is because modes with larger $k$ generally exit the horizon
later during inflation, allowing them to benefit more from the increasing
instability parameter $\xi$. 

If $k$ is very large, however, the amplification can be changed qualitatively
or even diminish. For $k_{{\rm ref}}<k<2|\xi|k_{{\rm ref}}$, the
mode never exits the horizon but the tachyonic instability, which
corresponds to $k^{2}-2\lambda\xi k{\cal H}<0$ in Eq.~\eqref{eq:-6},
can still occur. In this case, we expect that the amplification factor
can be even greater than  what is attainable with $k<k_{{\rm ref}}$.
If one further increases $k$ such that $k>2|\xi|k_{{\rm ref}}$,
the tachyonic instability cannot be reached because ${\cal H}$ cannot
exceed $k_{{\rm ref}}={\cal H}_{{\rm max}}$ and $k^{2}-2\lambda\xi k{\cal H}>k(k-2|\xi|k_{{\rm ref}})$
becomes always positive. In this case, the amplification factor is
expected to be suppressed. 

Indeed, these expectations are confirmed in our numerical solutions
of high-$k$ modes, as presented in the upper panel of Fig.~\ref{fig:high-k},
from which we see that $\sqrt{2k}|A_{+}|$  increases substantially
when $k$ increases from $0.01k_{{\rm ref}}$ (blue) to $0.5k_{{\rm ref}}$
(orange) and $6.0k_{{\rm ref}}$ (green). From $10k_{{\rm ref}}$
to $20k_{{\rm ref}}$, the amplification factor decreases with $k$. 

For $k<k_{{\rm ref}}$, it is possible for the mode to re-enter the
horizon during reheating, as is illustrated by the curve with $k=0.5k_{{\rm ref}}$.
Shortly after the end of inflation, this mode re-enters the horizon
and $A_{+}$ starts to evolve dynamically. The dip on this curve is
caused by $A_{+}$ passing zero. For higher-$k$ modes shown here,
similar dips also exist and become more frequent as $k$ increases. 

In the middle panel of Fig.~\ref{fig:high-k}, we show the evolution
of modes with the opposite helicity ($\lambda=-$). In the slow-roll
epoch, the opposite-helicity mode is either suppressed (for sub-horizon
evolution) or almost unaffected by $\xi$  (for super-horizon evolution),
as can be seen from this plot when $x\lesssim60$. However, their
post-inflationary evolution can be very dynamical if $k$ is comparable
or higher than $k_{{\rm ref}}$. This is because $\xi\propto\dot{\phi}$
often becomes negative during reheating (see the $\xi$ curve in the
lower panel of Fig.~\ref{fig:high-k}), leading to the amplification
of $A_{-}$ modes with large $k$.

\subsection{The Green function\label{subsec:The-Green-function}}

During slow-roll, the Green function can be approximately computed
using the analytical form ~\eqref{eq:-29}. In more general cases
including non-instantaneous reheating, the Green function has to rely
on numerical calculations. 

In general, the Green function can be constructed from solutions of
the corresponding homogeneous differential equation---see, e.g.,
Appendix B of Ref.~\cite{Domenech:2021ztg} for a brief introduction.
A simple and convenient way to construct it is given as follows~\cite{Garcia-Bellido:2023ser,vonEckardstein:2025oic}:
\begin{equation}
G_{k}\left(\eta,\tau\right)=\frac{{\rm Im}\left[u_{k}^{*}(\eta)u_{k}(\tau)\right]}{{\rm Im}\left[{u_{k}^{*}}'(\tau)u_{k}(\tau)\right]}\theta\left(\eta-\tau\right),\label{eq:-2}
\end{equation}
which is constructed using a solution of the following equation and
its conjugate:
\begin{equation}
u_{k}''+2{\cal H}u_{k}'+k^{2}u_{k}=0\thinspace.\label{eq:-20}
\end{equation}
Eq.~\eqref{eq:-20} is the mode equation of gravitons propagating
freely in the FLRW space. 

The most straightforward approach to compute the Green function is
to numerically solve Eq.~\eqref{eq:-20} and substituting the solution
into Eq.~\eqref{eq:-2}. However, we would like to raise a caveat
that numerical inaccuracies, albeit small, can be exponentially (due
to the exponential cosmic expansion during slow-roll) amplified by
large cancellations in this approach. We have observed this behavior
in numerical calculations and the analytical understanding of this
issue is presented in Appendix~\ref{sec:Green-F-App}. 

An alternative approach to obtain the Green function is to solve one
of the following equations:
\begin{align}
\left[\frac{\partial^{2}}{\partial\eta^{2}}+2{\cal H}(\eta)\frac{\partial}{\partial\eta}+k^{2}\right]G_{k}(\eta,\tau) & =\delta(\eta-\tau)\thinspace,\label{eq:-25}\\
\left[\frac{\partial^{2}}{\partial\tau^{2}}-2{\cal H}(\tau)\frac{\partial}{\partial\tau}-2{\cal H}'(\tau)+k^{2}\right]G_{k}(\eta,\tau) & =\delta(\eta-\tau)\thinspace,\label{eq:-26}
\end{align}
where ${\cal H}'$ is computed by ${\cal H}'=a''/a-(a'/a)^{2}=-\frac{a^{2}}{6}{\cal R}-{\cal H}^{2}$
with ${\cal R}$ the Ricci scalar determined by
\begin{equation}
{\cal R}=\frac{1}{M_{P}^{2}}\sum_{i}\left(3p_{i}-\rho_{i}\right).\label{eq:-27}
\end{equation}
Note that the $\tau$-based differential operator in front of $G_{k}$
in Eq.~\eqref{eq:-26} differs from the $\eta$-based differential
operator in Eq.~\eqref{eq:-25}.  They are adjoint operators of each
other. Although slightly more complicated than Eq.~\eqref{eq:-25},
Eq.~\eqref{eq:-26} offers a more convenient way to obtain the Green
function that can be directly used in Eq.~\eqref{eq:-18}, where $\tau$
varies from $-\infty$ to $\eta$ for each fixed value of $\eta$.
In our work, we compute $G_{k}(\eta,\tau)$ by evolving Eq.~\eqref{eq:-26}
backward starting at $\tau=\eta$, with the following initial condition:
\begin{equation}
\lim_{\tau\to\eta^{-}}G_{k}\left(\eta,\tau\right)=\eta-\tau\thinspace,\label{eq:-28}
\end{equation}
which can be proved using the Wronskian condition---see Appendix~\ref{sec:Green-F-App}
for further details. 

\begin{figure}
\centering

\includegraphics[width=0.95\textwidth]{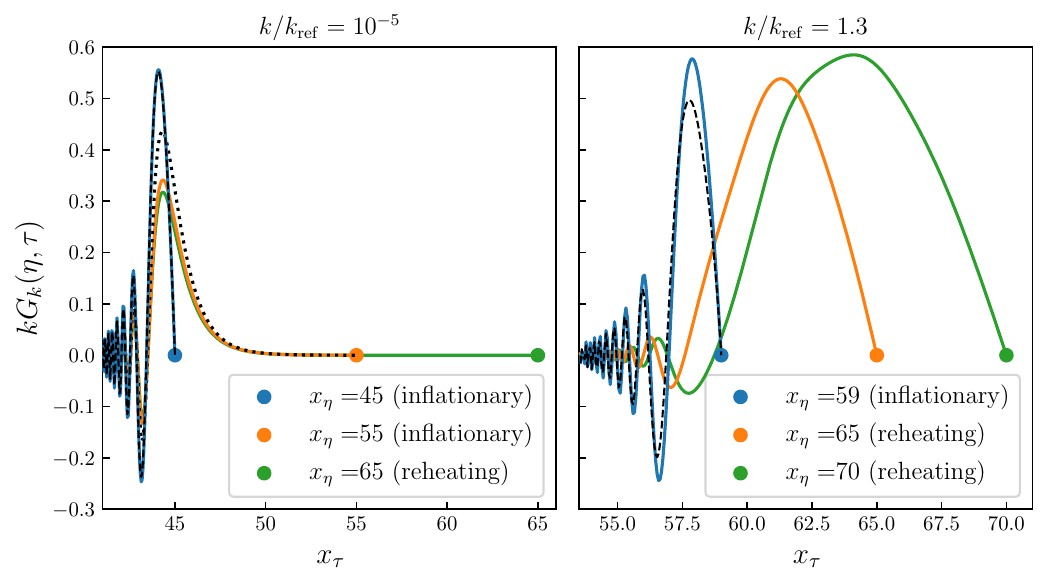}

\caption{Green functions for $k=10^{-5}k_{{\rm ref}}$ (left panel) and $k=1.3k_{{\rm ref}}$
(right panel) obtained by solving Eq.~\eqref{eq:-26} (solid lines),
compared with the analytical formula derived in de Sitter space (black
dashed or dotted lines)---see Eq.~\eqref{eq:-29}. In this figure,
$x_{\tau}$ and $x_{\eta}$ are defined as $x_{\tau}\equiv t(\tau)m_{\phi}$
and $x_{\eta}\equiv t(\eta)m_{\phi}$. Because the analytical formula
is valid only for $x_{\eta}\lesssim60$, we do not present the analytical
values when $x_{\eta}$ exceeds $60$. \label{fig:Green}}
\end{figure}

In Fig.~\ref{fig:Green}, we present two numerical examples of the
Green function with $k<k_{{\rm ref}}$ and $k>k_{{\rm ref}}$, obtained
by solving Eq.~\eqref{eq:-26} and compared with the analytical expression
in Eq.~\eqref{eq:-29} which only applies to de Sitter space. For
the background evolution we have used $\Gamma_{\phi}=0$ and $\Lambda^{-1}=11.2M_{P}^{-1}$.
In this figure, $x_{\tau}$ and $x_{\eta}$ are defined as $x_{\tau}\equiv t(\tau)m_{\phi}$
and $x_{\eta}\equiv t(\eta)m_{\phi}$,   and $x_{\eta}\approx60$
corresponds to the end of inflation. 

From Fig.~\ref{fig:Green}, we see that when $\eta$ is deep in the
slow-roll epoch,  in which it is a good approximation to assume de
Sitter space, the analytical Green function in Eq.~\eqref{eq:-29}
is in excellent agreement with the numerical result. This is demonstrated
 by the blue curve in comparison with  the black dashed curve in the
left panel. When $x_{\eta}$ is close to $60$, large deviations are
expected because the de Sitter approximation  becomes less accurate.
When $x_{\eta}$ exceeds $60$, Eq.~\eqref{eq:-29} becomes invalid
as it suffers from a divergence arising from the denominator when
$\tau$ passes zero. 

\subsection{Tensor power spectrum}

With the high-$k$ mode functions and the Green function numerically
available, we are now able to compute the tensor power spectrum generated
from inflationary and post-inflationary epochs. 

To proceed, we reformulate Eq.~\eqref{eq:-17} as follows
\begin{equation}
{\cal P}_{T,\lambda}^{{\rm ind}}\left(\eta,k\right)=\frac{k^{2}}{8\pi^{4}M_{P}^{4}}\int dpdq\sum_{\lambda_{1}\lambda_{2}}\left(1+\lambda\lambda_{1}c_{1}\right)^{2}\left(1+\lambda\lambda_{2}c_{2}\right)^{2}pq\left|I_{\tau}\right|^{2},\label{eq:-30}
\end{equation}
where we have used 
\begin{equation}
\int\frac{d^{3}\mathbf{p}}{(2\pi)^{3}}\to\int\frac{pq}{4\pi^{2}k}dpdq\thinspace,\label{eq:-31}
\end{equation}
which can be derived from the vector  triangle formed by $\mathbf{k}=\mathbf{p}+\mathbf{q}$
after taking a proper Jacobian into account. In Eq.~\eqref{eq:-30},
$c_{1}$ and $c_{2}$ are defined as $c_{1,2}=\cos\theta_{1,2}$ with
$\theta_{1,2}$ the angle between $\mathbf{k}$ and $\mathbf{p}$
or $\mathbf{q}$.  The advantage of formulating the integral in the
$p$-$q$ symmetric form is that the gauge field mode functions can
be pre-calculated on a series of discrete values of the momentum and
the sampling of $p$ and $q$ can be performed on a mesh constructed
from the discrete series. Note that for a given value of $k$, not
all samples on the mesh are kinematically allowed, as they must satisfy
\begin{equation}
|p-q|\leq k\leq p+q\thinspace.\label{eq:-32}
\end{equation}
If a sample satisfies Eq.~\eqref{eq:-32}, it is straightforward to
construct the triangle of $\mathbf{k}=\mathbf{p}+\mathbf{q}$ with
$c_{1}$ and $c_{2}$ determined by 
\begin{align}
c_{1} & =\frac{k^{2}+p^{2}-q^{2}}{2kp}\thinspace,\label{eq:-33}\\
c_{2} & =\frac{k^{2}+q^{2}-p^{2}}{2kq}\thinspace.\label{eq:-34}
\end{align}

With all the ingredients in Eq.~\eqref{eq:-30} ready for evaluation
on the $p$-$q$ mesh, the integral can be computed by summing over
all kinematically allowed samples, with $dp\to\Delta p$ and $dq\to\Delta q$,
where $\Delta p$ and $\Delta q$ denote the finite intervals of the
discrete momentum series. 

\begin{figure}
\centering

\includegraphics[width=0.98\textwidth]{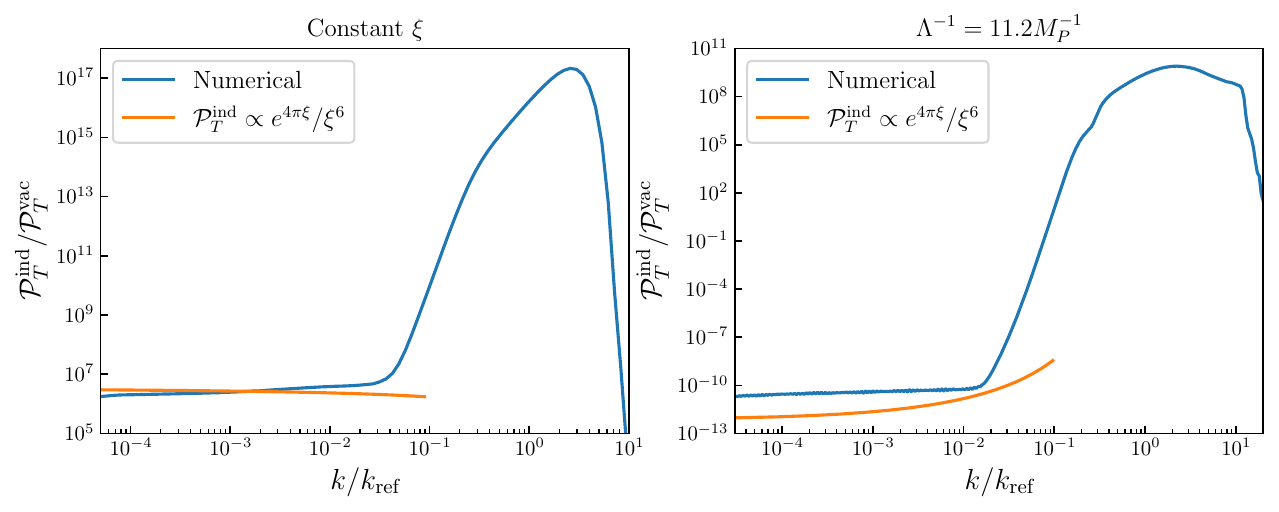}\caption{\label{fig:PT} Numerical and analytical values of ${\cal P}_{T}^{{\rm ind}}/{\cal P}_{T}^{{\rm vac}}$
for two cases: (i) $\xi$ is fixed at a constant ($\xi=5$) (ii) $\xi$
is determined by $\xi=\dot{\phi}/(2H\Lambda)$ and thus varies during
the evolution. }
\end{figure}

The numerical results of ${\cal P}_{T}^{{\rm ind}}$ calculated through
the above procedure are presented in Fig.~\ref{fig:PT} (blue lines),
together with the analytical estimates obtained using Eq.~\eqref{eq:-19}
(orange lines) for comparison. Similar to Fig.~\ref{fig:delta},
here we also make the comparison for two cases, one with constant
$\xi$ and the other with varying $\xi$ determined by $\xi=\dot{\phi}/(2H\Lambda)$.
For both cases, we have used $\Lambda^{-1}=11.2M_{P}^{-1}$ and $\Gamma_{\phi}=0.05m_{\phi}$.
In the first case, we see good agreement between the two curves for
$k\ll k_{{\rm ref}}$ (i.e., for modes that exit the horizon well
before the end of inflation). In the  second case, the difference
is significant. Given that the validity of Eq.~\eqref{eq:-19} requires
$\xi\gtrsim{\cal O}(1)$, which is not satisfied at early times of
the slow-roll epoch (see the lower panel of Fig.~\ref{fig:A-sol}),
and the rapid variation of $\xi$ at late times, we think such a large
difference is plausible. 

We are particularly interested in the peak of the this spectrum, which
can be estimated with a few crude approximations as follows.

First, the Green function makes its dominant contribution mainly via
the last peak shown in Fig.~\ref{fig:Green}. This peak is roughly
given by $G_{k}\sim\frac{1}{k}\sin k\delta_{\tau}$ with $\delta_{\tau}\equiv\eta-\tau$.
Focusing on this peak only, we estimate the $I_{\tau}$ integral in
Eq.~\eqref{eq:-18} as follows 
\begin{align}
I_{\tau} & \sim\int_{0}^{\pi/k}d\delta_{\tau}\frac{\sin k\delta_{\tau}}{ka^{2}(\tau)}pqA_{p}A{}_{q}\thinspace,\label{eq:-35}
\end{align}
where $A_{p}\equiv A_{\lambda_{1},p}(\tau)$ and $A_{q}\equiv A_{\lambda_{2},q}(\tau)$. 

During the reheating epoch, the variation of $a$ is relatively mild
and can be neglected in Eq.~\eqref{eq:-35}, as we only aims at an
order-of-magnitude estimate.  The high-$k$ gauge field modes are
enhanced by many orders of magnitude, as illustrated by Fig.~\ref{fig:high-k}.
For simplicity, we approximate them as
\begin{equation}
A_{\lambda,k}\sim\frac{1}{\sqrt{2k}}C_{k}\thinspace,\label{eq:-37}
\end{equation}
where $C_{k}$ is the large amplification factor which  can be obtained
from  Fig.~\ref{fig:high-k}. 

Putting the above pieces together, we obtain
\begin{equation}
I_{\tau}\sim\frac{\sqrt{pq}}{a^{2}k^{2}}C_{p}C_{q}\thinspace,\label{eq:-36}
\end{equation}
and hence 
\begin{equation}
{\cal P}_{T,\lambda}^{{\rm ind}}\left(\eta,k\right)\sim\frac{k^{2}}{8\pi^{4}M_{P}^{4}}\int dpdq(pq)\left|I_{\tau}\right|^{2}\sim\frac{C_{k}^{4}}{8\pi^{4}M_{P}^{4}}\frac{k^{4}}{a^{4}}\thinspace.\label{eq:-38}
\end{equation}
Taking a typical value of $k=6k_{{\rm ref}}$ for which $C_{k}\sim10^{6}$,
and using $a=\exp(2.0)\cdot a_{{\rm ref}}$ and $H(a_{{\rm ref}})=0.4H_{I}$
(both can be inferred from Fig.~\ref{fig:bkg}), we find ${\cal P}_{T,\lambda}^{{\rm ind}}/{\cal P}_{T}^{{\rm vac}}\sim10^{9}$,
which roughly agrees with the peak height in the right panel of Fig.~\ref{fig:PT}. 

\subsection{Revisit the conditions for the WB regime}

\begin{figure}
\centering

\includegraphics[width=0.98\textwidth]{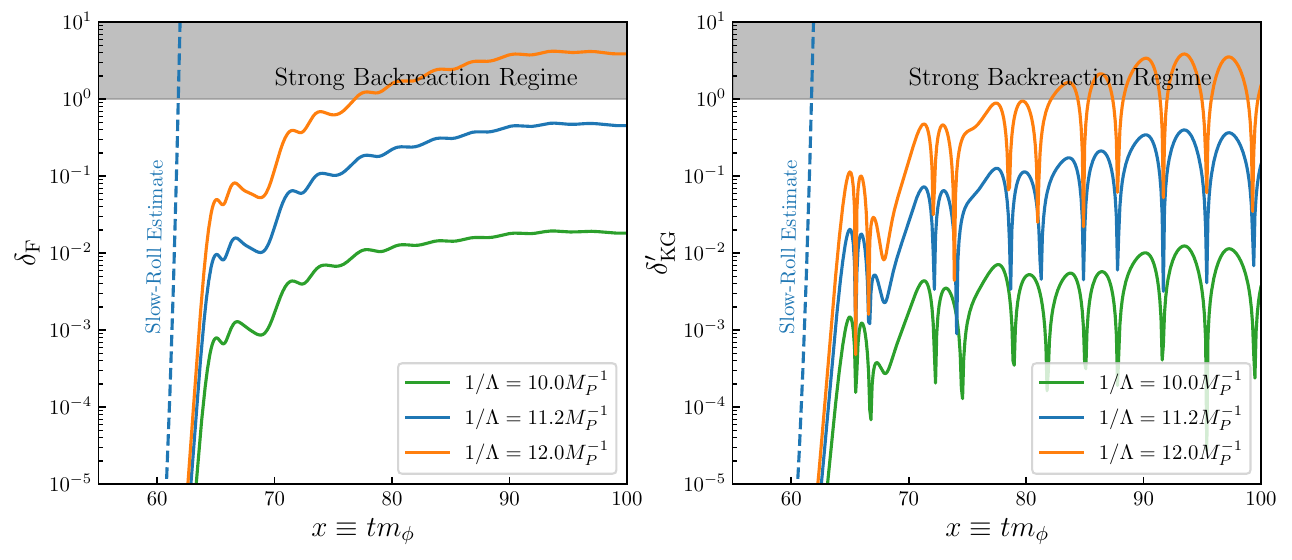}

\caption{Post-inflationary evolution of the two backreaction parameter, $\delta_{\text{F}}$
and $\delta'_{\text{KG}}$. \label{fig:delta-new}}
\end{figure}

In Fig.~\ref{fig:delta}, we have seen that the analytical estimates
of the two backreaction parameters, $\delta_{\text{F}}$ and $\delta_{\text{KG}}$,
become invalid when the evolution approaches the end of inflation,
because $\xi$ increases significantly and rapidly. The numerical
results shown in Fig.~\ref{fig:delta} suggest that a relatively
large $\xi$ (e.g., above $6$) occurring in a concrete inflationary
model may still be able to maintain $\delta_{\text{F}}\ll1$ and $\delta_{\text{KG}}\ll1$
around this epoch. However, since the numerical values of $\delta_{\text{F}}$
and $\delta_{\text{KG}}$ also grow rapidly at the end of inflation,
one may be concerned that $\delta_{\text{F}}$ or $\delta_{\text{KG}}$
may exceed unity in the subsequent evolution. 

In the left panel of Fig.~\ref{fig:delta-new}, we show the subsequent
evolution of $\delta_{\text{F}}$, which is obtained by summing over
the contributions of all $k$ modes to the gauge field energy density---see
Appendix~\ref{sec:rhoA} for technical details. The three solid curves
are produced for three different values of the axion-gauge coupling,
$1/\Lambda=10\thinspace M_{P}^{-1}$, $11.2\thinspace M_{P}^{-1}$,
and $12\thinspace M_{P}^{-1}$, assuming $\Gamma_{\phi}=0.1m_{\phi}$.
  For comparison, we also show the analytical estimate of $\delta_{\text{F}}$
(blue dashed) for $1/\Lambda=11.2\thinspace M_{P}^{-1}$. While the
analytical $\delta_{\text{F}}$ curve quickly enters the SB regime,
the actual growth of $\delta_{\text{F}}$ significantly slows down
in the post-inflationary evolution. Among the three selected values
of $1/\Lambda$, $10\thinspace M_{P}^{-1}$ and $11.2\thinspace M_{P}^{-1}$
remain in the WB regime, while $12\thinspace M_{P}^{-1}$ enters the
SB regime at $tm_{\phi}\approx77$. We note here that the heights
of these curves are highly sensitive to $\Lambda^{-1}$ because $\delta_{\text{F}}$
is exponentially sensitive to $\xi\propto\Lambda^{-1}$ at large $\xi$.

The other backreaction parameter, $\delta_{\text{KG}}$, requires
some modifications to make it applicable to the post-inflationary
evolution where the slow-roll approximation is no longer valid. The
original definition of $\delta_{\text{KG}}$ in Eq.~\eqref{eq:-22}
relies on the assumption that during slow-roll, the Hubble friction
term $3H\dot{\phi}$ approximately cancels the potential gradient
$dV/d\phi$. Consequently, one only needs to compare $\rho_{EB}/\Lambda$
with one of the two terms to quantify the impact of the gauge field
on the inflaton. During reheating, the two terms can have very different
magnitudes. In addition, $\Gamma_{\phi}$ in the $(3H+\Gamma_{\phi})\dot{\phi}$
term also starts to play a crucial role. Taking these into account,
we modify the definition of $\delta_{\text{KG}}$ as follows:
\begin{equation}
\delta'_{\text{KG}}\equiv\left|\frac{\rho_{EB}}{\Lambda X}\right|\ \ \ \text{with\ }X=\max\left\{ (3H+\Gamma_{\phi})\dot{\phi},\ dV/d\phi\right\} .\label{eq:-59}
\end{equation}
During slow-roll, $\delta'_{\text{KG}}$ reduces to $\delta{}_{\text{KG}}$
defined in Eq.~\eqref{eq:-22}. 

 In the right panel of Fig.~\ref{fig:delta-new}, we plot the post-inflationary
evolution of $\delta'_{\text{KG}}$, from which one can draw similar
conclusions as those from the left panel. The two plots of $\delta{}_{\text{F}}$
and $\delta'_{\text{KG}}$ suggest that once $\delta_{\text{F}}\ll1$
is satisfied, $\delta'_{\text{KG}}\ll1$ is also satisfied, and vice
versa. 

\section{High-frequency GWs\label{sec:hfgw}}

\begin{figure}
\centering

\includegraphics[width=0.6\textwidth]{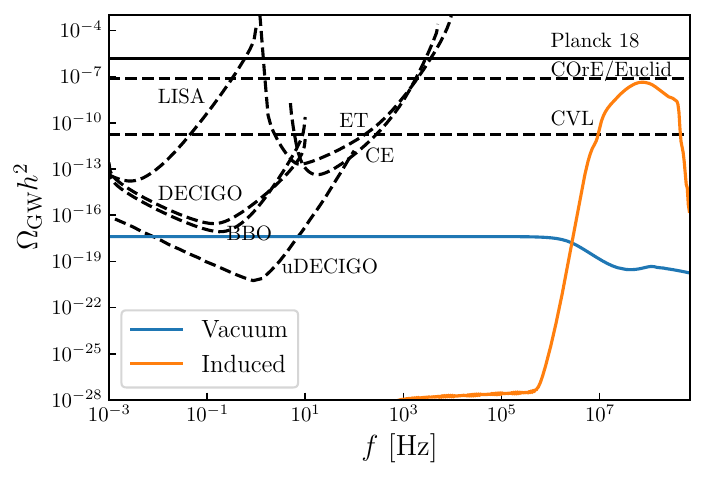}

\caption{GW power spectrum generated in axion inflation, including vacuum (blue)
and gauge-field-induced (orange) contributions. The projection curves
of GW interferometer experiments (LISA, BBO, etc.) are plotted using
data from \cite{Schmitz:2020syl} and \cite{Ringwald:2020ist}.
The horizontal black lines are derived from $N_{{\rm eff}}$ constrains.\label{fig:final-result}}
\end{figure}

As we have seen from the previous section, the gauge-field-induced
tensor power spectrum is substantially enhanced at the end of inflation
and during reheating. This implies that in the WB regime, axion inflation
naturally predicts the presence of high-frequency primordial GWs,
in contrast to usual inflationary scenarios in which the resulting
primordial GW spectrum at high frequencies is typically suppressed---see,
e.g., \cite{Mudrunka:2026kgm,Wang:2026ule,Wang:2026pff} or the blue
curve in Fig.~\ref{fig:final-result}. 

To recast the obtained tensor power spectrum into the GW energy-density
power spectrum, $\Omega_{{\rm GW}}$, we adopt the following formula~\cite{Saikawa:2018rcs,Boyle:2005se}:
\begin{equation}
\Omega_{{\rm GW}}(f)=\frac{\Omega_{\gamma}}{24}\left(\frac{g_{\star\rho}}{g_{\gamma}}\right)\left(\frac{g_{\star s,0}}{g_{\star s}}\right)^{\frac{4}{3}}{\cal P}_{T}(k)\thinspace,\label{eq:-60}
\end{equation}
where $\Omega_{\gamma}\approx2.473\times10^{-5}h^{2}$ is the present-day
photon energy density fraction, $g_{\gamma}=2$ accounts for the two
degrees of freedom of photons, and $g_{\star s,0}\approx2+\frac{7}{8}\times\frac{4}{11}\times6\approx3.9$
is the present value of $g_{\star s}$. Here and below, we use the
subscript ``0'' to  indicate present-day values. The GW frequency
$f$ is related to the physical momentum $k_{{\rm phy}}=k/a$ by $f=k_{{\rm phy},0}/(2\pi)$,
i.e., $f=k/(2\pi a_{0})$.  After reheating, the scale factor can
be determined from entropy conservation: $g_{\star s}(aT)^{3}=g_{\star s,0}(a_{0}T_{0})^{3}$.
 With $\rho_{R}=\frac{\pi^{2}}{30}g_{\star\rho}T^{4}$ and $T_{0}=2.73\ \text{K}$,
we obtain
\begin{equation}
f\approx43.1\ \text{GHz}\ g_{\star\rho}^{1/4}\left(\frac{g_{\star s,0}}{g_{\star s}}\right)^{\frac{1}{3}}\left(\frac{k_{{\rm phy}}}{\rho_{R}^{1/4}}\right)_{a\gg a_{{\rm rh}}},\label{eq:-61}
\end{equation}
where  $k_{{\rm phy}}/\rho_{R}^{1/4}$ approaches constant after
reheating as long as $g_{\star\rho}$ and $g_{\star s}$ do not vary.
It can be evaluated at any point well after reheating and before $g_{\star\rho}$
or $g_{\star s}$ starts varying. 

In Fig.~\ref{fig:final-result}, we present the obtained GW power
spectrum, including vacuum (blue) and gauge-field-induced (orange)
contributions. In this plot, we have used $1/\Lambda=11.2M_{P}^{-1}$
and $\Gamma_{\phi}=0.05m_{\phi}$.  At low frequencies, the vacuum
spectrum is nearly scale-invariant, a well-known feature predicted
by slow-roll inflation.  At sufficiently high frequencies with which
the graviton mode function evolves adiabatically throughout,  the
spectrum  should be suppressed~\cite{Mudrunka:2026kgm,Wang:2026pff}.
 The vacuum spectrum presented here is obtained using the code from~\cite{Wang:2026pff}.
Compared to the vacuum spectrum, the gauge-field-induced spectrum
 is orders of magnitude higher  at high frequencies from MHz to
GHz. This is due to the exponential amplification previously discussed,
which becomes particularly strong at the end of inflation and during
reheating. 

In Fig.~\ref{fig:final-result}, we also show the projection curves
of GW interferometer experiments including LISA~\cite{LISA:2017pwj},
the Big Bang Observer (BBO)~\cite{Crowder:2005nr,Corbin:2005ny},
the Cosmic Explorer (CE)~\cite{Reitze:2019iox}, the Einstein Telescope
(ET)~\cite{ET:2019dnz}, DECIGO~\cite{Seto:2001qf,Kudoh:2005as}
and ultimate-DECIGO (uDECIGO)~\cite{Kuroyanagi:2014qza}. Most of
these projection curves are plotted using data from Ref.~\cite{Schmitz:2020syl}
except for uDECIGO, which is taken from Ref.~\cite{Ringwald:2020ist}.
As is shown in Fig.~\ref{fig:final-result}, although some of these
experiments may be promising in probing the nearly scale-invariant
part of the vacuum spectrum, the most characteristic part---the substantially
enhanced high-frequency GW spectrum---lies far beyond the reach of
these experiments due to the over high frequencies. For high-frequency
GW detection, there has been a growing body of novel ideas, including
resonant cavities~\cite{Herman:2020wao,Herman:2022fau}, graviton-photon
conversion via the Gertsenshtein effect~\cite{Domcke:2020yzq}, 
Rydberg atoms~\cite{Kanno:2023whr}, and axion haloscopes~\cite{Domcke:2022rgu}---see
\cite{Aggarwal:2020olq,Aggarwal:2025noe} for reviews. However, to
our knowledge, the sensitivity reach of practical laboratory searches
is  still weaker than the cosmological constraint derived from  $N_{{\rm eff}}$.
The latest $N_{{\rm eff}}$ measurement from Planck 2018~\cite{Planck:2018vyg},
$N_{{\rm eff}}=2.99\pm0.34$ (2$\sigma$ C.L.), combined with $\Omega_{{\rm GW}}h^{2}\lesssim5.6\times10^{-6}\Delta N_{{\rm eff}}$~\cite{Caprini:2018mtu}
where $\Delta N_{{\rm eff}}=N_{{\rm eff}}-3.045$, implies an upper
bound of $\Omega_{{\rm GW}}h^{2}\lesssim1.6\times10^{-6}$, corresponding
to the horizontal black solid line in Fig.~\ref{fig:final-result}.
Future experiments, COrE~\cite{COrE:2011bfs} and Euclid~\cite{EUCLID:2011zbd},
are expected to improve the $N_{{\rm eff}}$ measurement to $\Delta N_{{\rm eff}}\lesssim0.013$
(2$\sigma$ C.L.), corresponding to $\Omega_{{\rm GW}}h^{2}\lesssim7.3\times10^{-8}$.
In addition, it has been reported~\cite{Ben-Dayan:2019gll} that
the cosmic-variance-limited (CVL) CMB polarization measurement may
be able to reach $\Delta N_{{\rm eff}}\lesssim3.1\times10^{-6}$,
corresponding to $\Omega_{{\rm GW}}h^{2}\lesssim1.7\times10^{-11}$.
These are presented in Fig.~\ref{fig:final-result} as horizontal
black dashed lines.  As is shown, the high-frequency GWs might be
relevant to future $N_{{\rm eff}}$ measurements. 

Finally, we would like to emphasize that, despite the lack of dedicated
GW experiments capable of probing the high-frequency spectrum, a careful
calculation in this regime is still important for several reasons.
First, it addresses the question of how the production of GWs in axion
inflation would be affected by non-instantaneous reheating. Second,
compared to various studies on high-frequency GWs produced during
reheating from, e.g., inflaton decay~\cite{Barman:2023ymn,Barman:2023rpg,Bernal:2023wus,Kanemura:2023pnv,Tokareva:2023mrt,Barman:2024htg,Jiang:2024akb,Das:2025cqs,Bernal:2026dsu}
or particle scattering/annihilation~\cite{Choi:2024ilx,Xu:2024fjl,Bernal:2024jim,Bernal:2025lxp,Xu:2025wjq},
our calculation reveals that axion inflation may give rise to one
of the strongest signals at high frequencies, thereby motivating novel
experimental developments in  high-frequency GW detection. Finally,
we note the possibility that if the subsequent evolution is not purely
radiation dominated, the high-frequency spectrum may be significantly
redshifted, potentially matching the frequency range accessible to
GW interferometer experiments.We leave this possibility for future
exploration. 

\section{Conclusion \label{sec:conclusion}}

In this work, we have revisited the dynamics of gauge field with tachyonic
enhancement and primordial gravitational wave (GW) production in axion
inflation. We restricted our analysis to the Weak Backreaction (WB)
regime to avoid the computational complexities and the potential risk
of overproducing scalar perturbations in the Strong Backreaction (SB)
regime. A crucial question we aimed to address is how the tachyonic
enhancement of the gauge field behaves when the evolution approaches
the end of inflation and subsequently enters reheating, given that
the enhancement is known to keep increasing during slow-roll. The
WB regime allows us to continuously track the evolution from slow-roll
through reheating to radiation domination, presenting a complete picture
of GW production in this framework. 

Our detailed numerical analysis revealed important limitations in
the analytical approximations used in previous studies. While known
analytical formulae for the tachyonic enhancement and for the gauge
field energy density hold well when the instability parameter $\xi$
varies sufficiently slowly,  we found significant discrepancies (see
the dashed lines in Figs.~\ref{fig:A-sol}, \ref{fig:delta}, and
\ref{fig:high-k}) when $\xi$ undergoes rapid variation---a generic
and inevitable feature near the end of inflation. Consequently, our
findings indicate that previous analytical formulae may lead to an
overestimate of the parameter space that falls into the SB regime,
emphasizing the necessity of precise numerical tracking during this
transitional phase. In addition, the analytical Green function used
for GW production during slow-roll may also become inaccurate near
the end of inflation, as illustrated in Fig.~\ref{fig:Green}. 

Although the WB regime of axion inflation is known to be incapable
of generating GW signals observable in proposed interferometer experiments,
it nevertheless leads to interesting and highly non-trivial phenomenological
consequences. During the reheating phase, the oscillatory behavior
of the inflaton leads to frequent sign-flips of $\xi$, exciting both
helical modes of the gauge field. The magnitude of $\xi$ during this
phase remains generally high compared to its typical slow-roll values.
Consequently,  axion inflation in the WB regime yields a high-frequency
GW signal that is many orders of magnitude stronger than the inflationary
production. The signal is sufficiently large to be relevant for future
precision measurements of the effective number of neutrino species,
$N_{{\rm eff}}$.   Our work places axion inflation among the most
efficient known primordial mechanisms for generating high-frequency
GWs, thereby providing a powerful theoretical motivation for the development
of novel high-frequency GW detectors.

\begin{acknowledgments}
This work is supported in part by the National Natural Science Foundation
of China under grant No.~12141501 and also by the CAS Project for
Young Scientists in Basic Research (YSBR-099). 
\end{acknowledgments}

\appendix

\section{Details of solving the mode equation}

When numerically solving the mode equation \eqref{eq:-6}, we parametrize
the mode function as follows:
\begin{equation}
A_{\lambda}=\frac{1+D}{\sqrt{2k}}e^{-ik\eta}\thinspace,\label{eq:-42}
\end{equation}
which we refer to as the $D$ parametrization. According to our experience,
it generally increases the numerical stability and accuracy compared
to the original form. With the $D$ parametrization, the initial conditions
become simple
\begin{equation}
\lim_{\eta\to-\infty}D(\eta)=0\thinspace,\ \ \lim_{\eta\to-\infty}D'(\eta)=0\thinspace.\label{eq:-49}
\end{equation}
 In practice, one has to solve the differential equation starting
from a finite value of $\eta$ instead of $\eta=-\infty$. In this
case, the initial conditions $D=D'=0$ at a finite $\eta$ may slightly
deviate from the true values but we find that the numerical inaccuracy
caused by this is significantly lower than that caused by directly
using Eq.~\eqref{eq:-9}.

With the $D$ parametrization, we rewrite the mode equation \eqref{eq:-6}
as follows:
\begin{equation}
D''-2ikD'-2\lambda\xi k{\cal H}\left(D+1\right)=0\thinspace.\label{eq:}
\end{equation}

It is important to notice that the conformal time $\eta$ may vary
over many orders of magnitude from slow-roll to reheating. Hence it
is not an appropriate time variable to be used in a numerical differential
equation solver. Throughout this work, we solve all differential equations
in physical time $t$ or its dimensionless proxy $x\equiv tm_{\phi}$.
 The advantage of using $t$ as the time variable is that it is approximately
linearly related to the number of e-fold during slow roll, and also
linearly related to the number of oscillations of $\phi$ during reheating. 

In physical time $t$, Eq.~\eqref{eq:-6} and Eq.~\eqref{eq:} become
\begin{align}
\ddot{A}_{\lambda}+H\dot{A_{\lambda}}+\left(\frac{k^{2}}{a^{2}}-2\lambda\xi\frac{k}{a}H\right)A_{\pm} & =0\thinspace,\label{eq:-48}\\
\ddot{D}+\left(H-\frac{2ik}{a}\right)\dot{D}-\frac{2\lambda\xi k}{a}H\left(1+D\right) & =0\thinspace.\label{eq:-50}
\end{align}
Defining $H_{x}\equiv H/m_{\phi}$, we further write Eq.~\eqref{eq:-50}
into the following dimensionless form:
\begin{equation}
\frac{d^{2}D}{dx^{2}}+\left(H_{x}-\frac{2ik}{m_{\phi}a}\right)\frac{dD}{dx}-\frac{2\lambda\xi k}{m_{\phi}a}H_{x}\left(1+D\right)=0\thinspace,\label{eq:-51}
\end{equation}
which is the final form used in our numerical calculation.

\section{Analytical discussion of the Green function \label{sec:Green-F-App}}

In this appendix, we present an analytical discussion of the Green
function relevant to our numerical analysis. 

First, the mode equation \eqref{eq:-20} can be reformulated as
\begin{equation}
\chi_{k}''+\left(k^{2}-\mu^{2}\right)\chi_{k}=0\thinspace,\label{eq:-62}
\end{equation}
with
\begin{align}
\chi_{k} & \equiv au_{k}\thinspace,\label{eq:-63}\\
\mu^{2} & \equiv\frac{a''}{a}=-\frac{a^{2}}{6}{\cal R}\thinspace,\label{eq:-64}
\end{align}
where the Ricci scalar ${\cal R}$ can be computed using Eq.~\eqref{eq:-27}.
If ${\cal R}$ has a simple form, Eq.~\eqref{eq:-62} can often be
solved analytically. This includes scenarios when  the universe is
dominated by vacuum energy, matter or radiation---see, e.g., Eqs.~(3.1)-(3.4)
of Ref.~\cite{Wang:2026pff}. For instance, during slow-roll where
${\cal R}=-12H^{2}$ is approximately a constant, the solution reads
\begin{equation}
\chi_{k}=\frac{1}{\sqrt{2k}}\left(1-\frac{i}{k\eta}\right)e^{-ik\eta}\thinspace.\label{eq:-65}
\end{equation}

In canonical quantization, the mode function needs to be properly
normalized. The normalization gives rise to the Wronskian condition
\begin{equation}
\chi_{k}{\chi_{k}^{*}}'-{\chi_{k}^{*}}\chi_{k}'=i\thinspace.\label{eq:-66}
\end{equation}

With the Wronskian condition, the denominator of Eq.~\eqref{eq:-2}
can be reduced to 
\begin{equation}
{\rm Im}\left[{u_{k}^{*}}'(\tau)u_{k}(\tau)\right]=\frac{1}{a(\tau)^{2}}{\rm Im}\left[{\chi_{k}^{*}}'(\tau)\chi_{k}(\tau)\right]=\frac{1}{2a(\tau)^{2}}\thinspace.\label{eq:-67}
\end{equation}
Correspondingly, Eq.~\eqref{eq:-2} simplifies to 
\begin{equation}
G_{k}\left(\eta,\tau\right)=2a(\tau)^{2}{\rm Im}\left[u_{k}^{*}(\eta)u_{k}(\tau)\right]\theta\left(\eta-\tau\right).\label{eq:-2-1}
\end{equation}

The Wronskian condition also has an important implication for the
numerator of Eq.~\eqref{eq:-2}. When $\tau$ approaches $\eta$,
one can expand it in terms of $\delta_{\tau}\equiv\eta-\tau$: 
\begin{align}
{\rm Im}\left[u_{k}^{*}(\eta)u_{k}(\tau)\right] & ={\rm Im}\left\{ u_{k}^{*}(\eta)\left[u_{k}(\eta)-u_{k}'(\eta)\delta_{\tau}+{\cal O}\left(\delta_{\tau}^{2}\right)\right]\right\} \nonumber \\
 & ={\rm Im}\left[-u_{k}^{*}(\eta)u_{k}'(\eta)\right]\delta_{\tau}+{\cal O}\left(\delta_{\tau}^{2}\right)\nonumber \\
 & =\frac{1}{2a(\eta)^{2}}\delta_{\tau}+{\cal O}\left(\delta_{\tau}^{2}\right).\label{eq:-68}
\end{align}
Hence we obtain the following asymptotic behavior of the Green function:
\begin{equation}
\lim_{\tau\to\eta^{-}}G_{k}\left(\eta,\tau\right)=\eta-\tau\thinspace.\label{eq:-69}
\end{equation}

Although Eq.~\eqref{eq:-2} or Eq.~\eqref{eq:-2-1} is conceptually
simple for computing the Green function, it may cause a numerical
problem in the presence of small inaccuracies of $u_{k}$ or $\chi_{k}$.
 Consider a quasi-de-Sitter background and the following form of
$\chi_{k}$:
\begin{equation}
\chi_{k}=\frac{1}{\sqrt{2k}}\left[1-\frac{i}{k\eta}\left(1+\epsilon\right)\right]e^{-ik\eta},\label{eq:-70}
\end{equation}
where $\epsilon\ll1$ accounts for a small deviation from the solution
in Eq.~\eqref{eq:-65}. Using Eq.~\eqref{eq:-70} and Eq.~\eqref{eq:-2-1}
to compute the Green function, we obtain
\begin{equation}
G_{k}\approx\left(1+\frac{2{\rm Im}\epsilon}{k\eta}+\frac{{\rm Re}\epsilon}{k^{2}\eta^{2}}+\frac{|\epsilon|^{2}}{k^{2}\eta^{2}}\right)\delta_{\tau}+{\cal O}\left(\delta_{\tau}^{2}\right).\label{eq:-71}
\end{equation}
Note that here we have only expanded $G_{k}$ in $\delta_{\tau}$
in order to inspect its behavior at small $\delta_{\tau}$, and no
expansion in $\epsilon$ has been performed. 

If $\epsilon$ satisfies 
\begin{equation}
|\epsilon|^{2}+{\rm Re}\epsilon+2k\eta{\rm Im}\epsilon=0\thinspace,\label{eq:-72}
\end{equation}
then Eq.~\eqref{eq:-71} would reproduce exactly the limit in Eq.~\eqref{eq:-69}.
In fact, it is straightforward to prove that if $\chi_{k}$ satisfies
the Wronskian condition in Eq.~\eqref{eq:-66}, then $\epsilon$ must
satisfy Eq.~\eqref{eq:-72}.

Numerically, however, the three $\epsilon$ term in Eq.~\eqref{eq:-71}
are particularly dangerous because numerical inaccuracies can easily
violate Eq.~\eqref{eq:-72}, implying that the three terms may fail
to fully cancel out each other. Because $1/|\eta|\approx aH$ grows
exponentially during slow-roll, numerical inaccuracies can be exponentially
amplified, causing catastrophic errors in the Green function. In practice,
we recommend the approach based on Eq.~\eqref{eq:-26} instead of
the more straightforward one based on Eq.~\eqref{eq:-2} for computing
the Green function.

\section{Calculation of $\rho_{A}$ and $\rho_{EB}$\label{sec:rhoA}}

The energy density of the gauge field, $\rho_{A}$, can be decomposed
into $EE$ and $BB$ contributions (see, e.g., Ref.~\cite{Jimenez:2017cdr}):
\begin{equation}
\rho_{A}=\rho_{EE}+\rho_{BB}\thinspace,\label{eq:-39}
\end{equation}
where 
\begin{align}
\rho_{EE} & =\frac{1}{2}\langle\mathbf{E}\cdot\mathbf{E}\rangle=\frac{1}{2a^{4}}\sum_{\lambda=\pm}\int\frac{d^{3}\mathbf{k}}{(2\pi)^{3}}\left|\frac{\partial}{\partial\eta}A_{\lambda}\left(k\right)\right|^{2},\label{eq:-40}\\
\rho_{BB} & =\frac{1}{2}\langle\mathbf{B}\cdot\mathbf{B}\rangle=\frac{1}{2a^{4}}\sum_{\lambda=\pm}\int\frac{d^{3}\mathbf{k}}{(2\pi)^{3}}k^{2}\left|A_{\lambda}\left(k\right)\right|^{2}.\label{eq:-41}
\end{align}
Similarly, $\rho_{EB}$ can also be expressed as an integral of the
mode functions:
\begin{equation}
\rho_{EB}=\langle\mathbf{E}\cdot\mathbf{B}\rangle=-\frac{1}{2a^{4}}\sum_{\lambda=\pm}\int\frac{d^{3}\mathbf{k}}{(2\pi)^{3}}k\frac{\partial}{\partial\eta}\left|A_{\lambda}\left(k\right)\right|^{2}.\label{eq:-47}
\end{equation}

If one naively performs the integration over $k$, Eqs.~\eqref{eq:-40}
and \eqref{eq:-41} are UV divergent.  Actually they are divergent
even in the Minkowski vacuum, due to vacuum fluctuations of the gauge
field. In de Sitter space, this is also divergent and the divergence
can be readily absorbed by the overall vacuum energy that is driving
inflation. In fact, the quantized inflaton field also introduces a
similar UV divergence in the energy density. Altogether, these UV
divergences can be absorbed by a shift of the potential energy $V$.

In this work, we are only concerned with the dynamical contributions
to $\rho_{EE}$ and $\rho_{BB}$, i.e., the contributions generated
by the $\phi F^{\mu\nu}\tilde{F}_{\mu\nu}$ interaction. Hence we
can subtract the vacuum contribution and only integrate over those
modes that have been excited by the interaction. Numerically, this
is implemented by setting a cut, $|1+D|>1+\epsilon$, where $\epsilon$
is a small number, and summing over the modes that meet this requirement.
We find that the results are broadly insensitive to the choice of
$\epsilon$ as long as the amplification caused by $\xi$ is significant.

Our numerical calculation of $\rho_{EE}$, $\rho_{BB}$, and $\rho_{EB}$
is performed also based on the $D$ parametrization {[}see Eq.~\eqref{eq:-42}{]}.
More specifically, they are evaluated as follows :
\begin{equation}
\rho_{X}=\int d\ln k\frac{d\rho_{X}}{d\ln k}\ \ \text{for}\ \ X\in\{EE,BB,EB\}\thinspace,\label{eq:-43}
\end{equation}
with
\begin{align}
\frac{d\rho_{EE}}{d\ln k} & \equiv\frac{1}{2a^{4}}\sum_{\lambda=\pm}\frac{k^{4}}{(2\pi)^{2}}\left[\left|1+D+\frac{D'}{-ik}\right|^{2}-1\right],\label{eq:-44}\\
\frac{d\rho_{BB}}{d\ln k} & \equiv\frac{1}{2a^{4}}\sum_{\lambda=\pm}\frac{k^{4}}{(2\pi)^{2}}\left[\left|1+D\right|^{2}-1\right],\label{eq:-45}\\
\frac{d\rho_{EB}}{d\ln k} & \equiv-\frac{1}{2a^{4}}\sum_{\lambda=\pm}\frac{k^{3}}{(2\pi)^{2}}2\Re\left[D'\left(1+D^{*}\right)\right].\label{eq:-46}
\end{align}

\bibliographystyle{JHEP}
\bibliography{ref}

\end{document}